\documentclass[10pt,journal,compsoc]{IEEEtran}
\pdfoutput=1
\usepackage[pdftex]{graphicx}
\usepackage[nocompress]{cite}
\usepackage[cmex10]{amsmath}
\usepackage{enumerate}
\usepackage{algorithmic}
\usepackage{array}
\usepackage{eqparbox}
\usepackage{url}
\usepackage{tikz}
\usetikzlibrary{arrows,fit,positioning,shapes,chains}

\pgfdeclarelayer{background}
\pgfsetlayers{background,main}

\begin{document}

\title{Breaking the Target: An Analysis of \\
Target Data Breach and Lessons Learned}

\author{Xiaokui~Shu,
        Ke~Tian*,
        Andrew~Ciambrone*
        and~Danfeng~(Daphne)~Yao,~\IEEEmembership{Member,~IEEE}
\IEEEcompsocitemizethanks{
    \IEEEcompsocthanksitem X. Shu, K. Tian, A. Ciambrone and D. Yao are with the
    Department of Computer Science, Virginia Tech, Blacksburg, VA,
    24060.\protect\\
    E-mail: \{subx, ketian, andrjc4, danfeng\}@vt.edu.
    \IEEEcompsocthanksitem *K. Tian and A. Ciambrone contribute equally to the paper.}
\thanks{}}

%

\IEEEcompsoctitleabstractindextext{
\begin{abstract}

This paper investigates and examines the events leading up to the second most
devastating data breach in history: the attack on the Target Corporation. It
includes a thorough step-by-step analysis of this attack and a comprehensive
anatomy of the malware named BlackPOS. Also, this paper provides insight into
the legal aspect of cybercrimes, along with a prosecution and sentence example
of the well-known TJX case. Furthermore, we point out an urgent need for
improving security mechanisms in existing systems of merchants and propose
three security guidelines and defenses. Credit card security is discussed at
the end of the paper with several best practices given to customers to hide
their card information in purchase transactions.

\end{abstract}

\begin{IEEEkeywords}
Data breach, information leak, point-of-sale malware, cybercrime, network
segmentation, security alert, system integrity, credit card security, EMV,
tokenization
\end{IEEEkeywords}}

\maketitle

\IEEEdisplaynotcompsoctitleabstractindextext
\IEEEpeerreviewmaketitle

\section{Introduction}

Between November 27 and December 18, 2013, the Target Corporation's network was
breached, which became the second largest credit and debit card breach after
the TJX breach in 2007. In the Target incident, 40 million credit and debit
card numbers and 70 million records of personal information were stolen.  The
ordeal cost credit card unions over two hundred million dollars for just
reissuing cards.

Target Corp. is not the only target of data breaches. Up to the 23rd of
September, 568 data breaches are reported in the year
2014~\cite{itrcbreachreport}. The latest significant breach, i.e., the Home
Depot breach, came to light in September 2014. As of September 14, it is known
that 23 out of 28 Home Depot stores in the State of Alabama were
breached~\cite{homedepotalabama}. The entire plot could involve a large portion
of the 2,200 Home Depot stores in the states and 287 stores overseas, which
might result in a larger breach than the Target breach. We list four other
significant breaches in the last two years. The increasing number and scale of
data breach incidents are alarming.

\begin{itemize}

    \item Sally Beauty Supply discovered in March 2014 that 282,000 cards were
        stolen~\cite{sallybreach}.

    \item Neiman Marcus reported that 1.1 million cards were stolen during July
        to October, 2013~\cite{neimanmarcus}.

    \item Michaels and Aaron Brother reported that 3 million cards were stolen
        from May 2013 to January 2014~\cite{MABbreaches}.

    \item P.F. Chang's data breach occurred from September 2013 to June 2014
        impacting over 7 million cards~\cite{changbreach}.

\end{itemize}

Securing massive amounts of connected systems is known to be technically
challenging, especially for retailers those possess vast networks across the
nation, like Target and Home Depot. Target security division attempted to
protect their systems and networks against cyber threats such as malware and
data exfiltration. Six months prior to the breach, Target deployed a well-known
and reputable intrusion and malware detection service named
FireEye~\cite{targetfireeye}, which was guided by the CIA during its early
development~\cite{fireeye}. Unfortunately, multiple malware alerts were
ignored. Some prevention functionalities were turned off by the administrators
who were not familiar with the FireEye system. Target Corp. missed the early
discovery of the breach.

This paper analyzes Target's data breach incident from both technical and legal
perspectives. The description of the incident and the analysis of the involved
malware explain how flaws in the Target's network were exploited and why the
breach was undiscovered for weeks. The Target data breach is still under
investigation and there is no arrest made known to the public. Even if the
perpetrators are identified, cyber crimes involving extradition are notorious
to prosecute. We discuss the difficulties of data breach discovery,
investigation and prosecution with respect to legislation and international
cooperation. An earlier incident, TJX data breach in 2007, is presented as the
precedent for arresting and sentencing criminals committing financial
cybercrimes.

\begin{figure*}
    \begin{center}
        \tikzstyle{anchornode}=[anchor=north, font=\footnotesize, 
    outer sep=5ex, text width=12ex]

\begin{tikzpicture}[font=\rmfamily\small,
    start chain=going base right,
    node distance=3.1mm]

  \node[on chain] (e1) {September~~~~};
  \node[anchornode, xshift=0.4ex] (t1) at (e1)
  {Attackers compromised Fazio Mechanical Services.};

  \node[on chain] (e2) {November 15};
  \node[anchornode, xshift=0.3ex] (t2) at (e2)
  {Attackers broke into Target's network and tested malware on POS machines.};

  \node[on chain] (e3) {November 27};
  \node[anchornode, xshift=0.3ex] (t3) at (e3)
  {Attackers began to collect credit card data.};

  \node[on chain] (e4) {November 30};
  \node[anchornode, xshift=0.2ex] (t4) at (e4)
  {POS malware fully installed.};
  \node[anchornode, outer sep=4ex] (t41) at (t4)
  {Attackers installed data exfiltration malware.};
  \node[anchornode, outer sep=6ex] (t42) at (t41)
  {Symantec and FireEye alerts triggered.};

  \node[on chain] (e5) {December 2};
  \node[anchornode, xshift=1ex] (t5) at (e5)
  {Attackers began to move credit card data out.};
  \node[anchornode, outer sep=6ex] (t51) at (t5)
  {Additional FireEye alerts triggered.};

  \node[on chain] (e6) {December 12};
  \node[anchornode, xshift=0.4ex] (t6) at (e6)
  {Department of Justice notified Target.};

  \node[on chain] (e7) {December 15};
  \node[anchornode, xshift=0.4ex] (t7) at (e7)
  {Target removed most malware.};

  \node[on chain] (f) {};

\definecolor{arrowcolor}{RGB}{220,110,120}

\begin{pgfonlayer}{background}
\node[
    inner sep=10pt,
    single arrow,
    single arrow head extend=0.8cm,
    draw=none,
    fill=arrowcolor,
    fit= (e1) (f)
] (arrow) {};
\end{pgfonlayer}

\end{tikzpicture}
        \caption{Timeline of the Target data breach (2013).}
        \label{fig:timeline}
    \end{center}
\end{figure*}
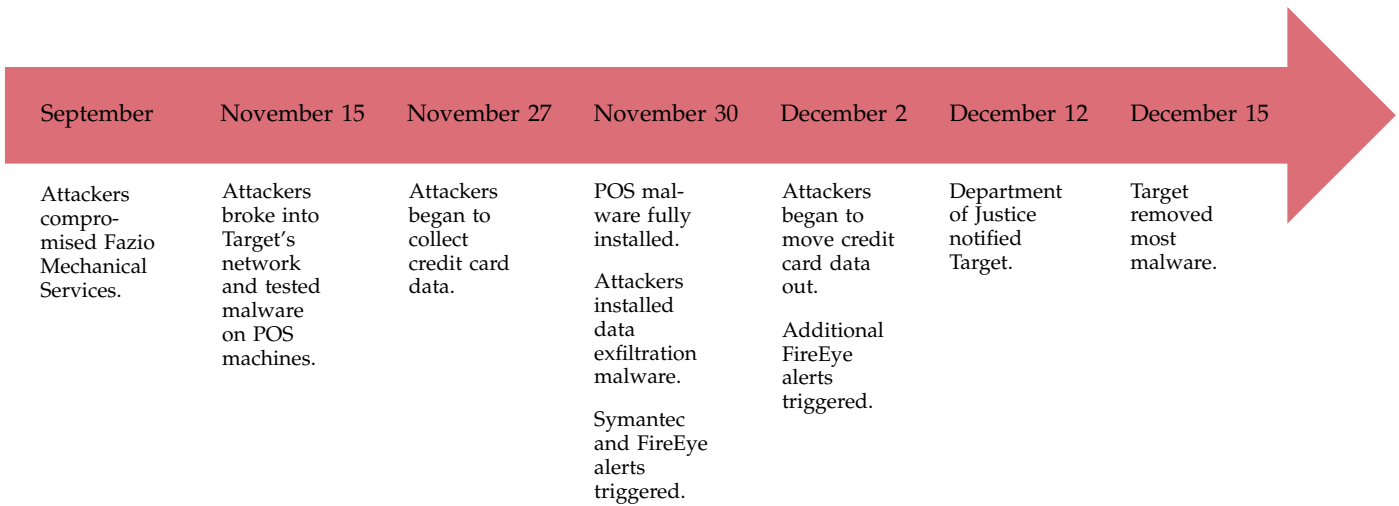

As we observe an increasing number of data breaches, these incidents bring us
to rethink the effectiveness of existing security mechanisms, solutions,
deployments and executions. Credit card breach has a huge negative impact on
every entity in the payment ecosystem, including merchants, banks, card
associations and customers. In this paper, we provide several insights into
weak links in the payment ecosystem, specifically in existing security
techniques and practices.
We give several best practice suggestions for merchants and customers to
enforce their data security and to minimize information leak.

The contributions of our work are summarized as follows.

\begin{itemize}

    \item We gather and verify information from multiple sources and describe
        the process of the Target data breach in details
        (Section~\ref{sec:incident}).

    \item We provide an in-depth analysis of the major malware used in the
        Target breach, including its design features for circumventing
        detections as well as the marketing of the malware
        (Section~\ref{sec:blackpos}).

    \item We discuss the complexities and challenges in data breach
        investigation and criminal prosecution, specifically from the legal
        perspective. We describe the TJX breach in 2007 as a precedent for
        arresting and sentencing cyber criminals (Section~\ref{sec:law}).

    \item We provide three security guidelines for merchants to enhance their
        payment system security: \textit{i)} payment system integrity
        enforcement, \textit{ii)} effective alert system design, and
        \textit{iii)} proper network segmentation
        (Section~\ref{sec:companysolutions}).

    \item We discuss the current status of credit card security, point out
        problems in the credit card system, and give customers best practices
        to hide their information in purchase transactions
        (Section~\ref{sec:customersolutions}).

\end{itemize}

\section{The Target Incident}
\label{sec:incident}

The systems and networks of Target Corp. were breached in November and
December, 2013, which results in 40 million card numbers and 70 million
personal records stolen~\cite{targetnumber}. Multiple parties get involved in
the federal investigation of the incident. The list includes United State
Secret Service, iSIGHT Partners, DELL SecureWorks, Seculert, the FBI, etc. In
addition, companies like HP, McAfee and IntelCrawler provide analysis of the
discovered malware, i.e., BlackPOS, and the marketing of the stolen cards.

\subsection{Breach Into Target}

There are multiple theories on how the criminals initially hacked into Target,
and none of them have yet been confirmed by Target Corporation. However, the
primary and most well-supported theory is that the initial breach didn't
actually occur inside Target~\cite{targetalarm}. Instead, it occurred in a
third party vendor, Fazio Mechanical Services, which is a heating, ventilation,
and air-conditioning firm.

\begin{figure*}
    \begin{center}
        \includegraphics[width=7.2in]{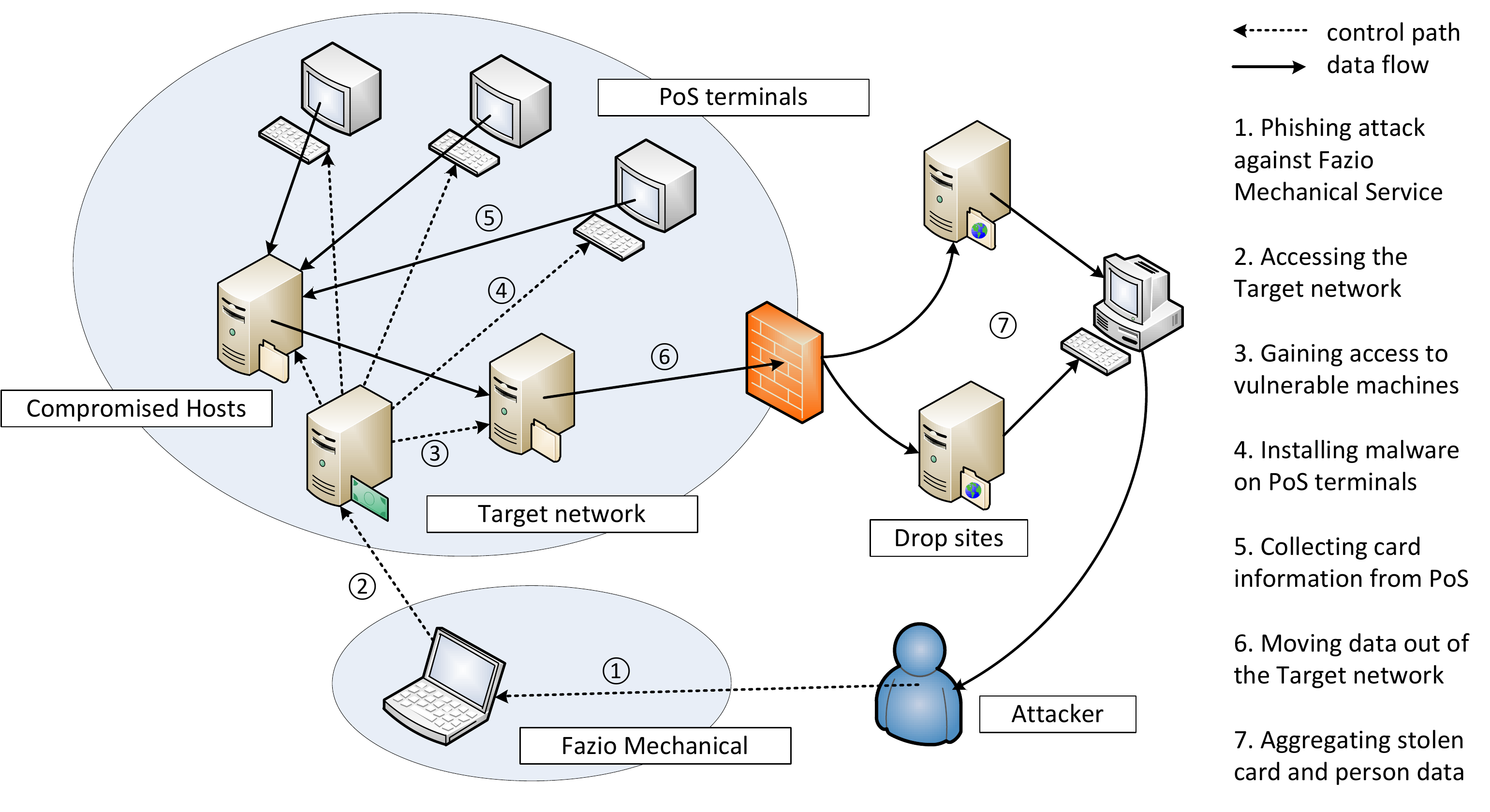}
        \caption{Attack steps of the Target breach.}
        \label{fig:targetbreachflow}
    \end{center}
\end{figure*}

According to this theory, we present the timeline of the incident in
Fig.~\ref{fig:timeline} and steps of the plot in
Fig.~\ref{fig:targetbreachflow}. Attackers first penetrated into the Target
network with compromised credentials from Fazio Mechanical. Then they probed
the Target network and pinpointed weak points to exploit. Some vulnerabilities
were used to gain access to the sensitive data, and others were used to build
the bridge transferring data out of Target. Due to the weak segmentation
between non-sensitive and sensitive networks inside Target, the attackers
accessed the point of sale networks.

\subsubsection{Phase I: Initial Infection}

At some point the Fazio Mechanical Services system was compromised by what is
believed to be a Citadel Trojan~\cite{targetphishingattack}. This Trojan was
initially installed through a phishing attempt. Due to the poor security
training and security system of the third party, the Trojan gave the attackers
full range of power over the company's system~\cite{targetalarm}. It is not
known if Fazio Mechanical Services was targeted, or if it was part of a larger
phishing attack to which it just happened to fall victim. But it is certain
that Fazio Mechanical had access to Target's Ariba external billing system, or
the business section of Target network.

\subsubsection{Phase II: PoS Infection}

Due to Target's poor segmentation of its network, all that the attackers needed
in order to gain access into Target's entire system was to access its business
section. From there, they gained access to other parts of the Target network,
including parts of the network that contained sensitive data. Once they gained
access into Target's network they started to test installing malware onto the
point of sales devices. The attackers used a form of point of sales malware
called BlackPOS, which is further discussed in Section~\ref{sec:blackpos}.

\subsubsection{Phase III: Data collection}

Once BlackPOS was installed, updated and tested. The malware started to scan
the memory of the point of sales to read the track information, especially card
numbers, of the cards that are scanned by the card readers connected to the
point of sales devices.

\subsubsection{Phase IV: Data exfiltration}

The card numbers were then encrypted and moved from the point of sales devices
to internal repositories, which were compromised machines. During the breach
the attackers took over three FTP servers on Target's internal network and
carefully chose backdoor user name ``Best1\_user'' with password
``BackupU\$r'', which are normally created by IT management software
\textit{Performance Assurance for Microsoft Servers}. During peak times of the
day, the malware on the point of sale devices would send credit card
information in bulk to the closest FTP Server~\cite{hackingpos}. The stolen
card information is then relayed to other compromised machines and finally
pushed to drop sites in Miami and Brazil~\cite{dellsecureworks}.

\subsubsection{Phase V: Monetization}

Sources indicate the stolen credit card information was aggregated at a server
in Russia, and the attackers collected 11~GB data during November and December
2013. The credit cards from the Target breach were identified on black market
forums for sell~\cite{sellingcreditcard}. At this point, it is unclear how
these sellers, e.g., Rescator (nick name), is connected with the stolen card
and personal information. In Section~\ref{tjxbreach}, we describe the well
studied case of TJX credit card breach. It hints possible paths of peddling
stolen credit cards in the black market.




\subsection{Targets Security}
\label{sec:incident:security}

Target did not run their systems and networks without security measures. They
had firewalls in place and they attempted to segment their network using
Virtual local area networks (VLAN)~\cite{targetfireeye}. Target also deployed
FireEye, a well-known network security system, six months prior to the breach.
FireEye provides multiple levels of security from malware detection to network
intrusion detection system (NIDS). 

However, the breach demonstrates that sensitive data in Target, e.g., credit
card information and personal records, is far from secure. Target failed at
detecting or preventing the breach at several points, among which we list the
four most vital ones:

\begin{itemize}

    \item Target did not investigate into the security warnings generated by
        multiple security tools, e.g., FireEye, Symantec, and certain malware
        auto-removal functionalities were turned off~\cite{targetwarningsigns}.

    \item Target did not take correct methods to segment their systems, failing
        to isolate their sensitive network assets from easily accessed network
        sections. The VLAN technique used for segmentation is reported easy to
        get around~\cite{forrester}.

    \item Target did not harden their point of sale terminals, allowing
        unauthorized software installation and configuration. The settings
        resulted in the spread of malware and sensitive card information read
        from point of sale terminals.
        
    \item Target did not apply proper access control on verities of accounts
        and groups, especially the ones from third party
        partners~\cite{targetemail}. The failure resulted in the initial
        break-in from the HVAC company Fazio Mechanical Services Inc.
        

\end{itemize}

\subsection{After the Breach}

The former CEO of the company, Gregg Steinhafel, resigned after the breach.
Target appointed a new chief information officer Bob DeRodes and provided
details on enhancing their security with 100 million
dollars~\cite{targetstatement}. The plan includes upgrading insecure point of
sale machines and deploying chip-and-PIN-enabled technology for payment.
Defenses such as better segmentation of the network, comprehensive log analysis
and stricter access control are also mentioned in the plan.

\section{BlackPOS}
\label{sec:blackpos}

BlackPOS, seen on underground forums since February
2013~\cite{posmalwaredates}, is believed to be the major malware used in the
data breaches at Target (2013), P.F. Chang's (2013), and Home Depot (2014). The
malware is a form of memory scrapper that takes a chunk of a systems memory and
looks for credit card numbers. We describe the functionalities of BlackPOS
captured in the Target breach, discuss its design features for circumventing
detection techniques, and present the investigations of POS malware development
and marketing.

\begin{figure}
    \begin{center}
        \begin{tikzpicture}[
    font=\rmfamily\small,
    every matrix/.style={column sep=1.06cm,row sep=3ex},
    source/.style={draw,rounded corners,fill=yellow!20,
    minimum height=4ex, minimum width=16ex},
    normalnode/.style={source,fill=white!20},
    every node/.style={align=center}]

    \matrix{
        \node[source] (s) {register service};
        & \node[source] (p) {scan process list}; \\

        \node[normalnode] (ss) {start service};
        & \node[normalnode] (pp) {select process}; \\

        & \node[normalnode] (ppp) {scan process}; \\
        
        \node[source] (u) {check time};
        & \node[normalnode] (pppp) {scan mem chunks}; \\

        \node[normalnode] (uu) {upload log};
        & \node[normalnode] (ppppp) {extract track info}; \\
    };

    \draw[-latex] (s) -- (ss);
    \draw[-latex] (p) -- (pp);
    \draw[-latex] (pp) -- (ppp);
    \draw[-latex] (ppp) -- (pppp);
    \draw[-latex] (pppp) -- (ppppp);
    \draw[-latex] (u) -- (uu);

    \node[draw,dashed,fit=(s) (ss),inner sep=2ex, yshift=2.4ex,
    minimum height=3.2cm] (sa) {};
    \node[above=-5.4ex of sa] {program \\ maintenance};

    \node[draw,dashed,fit=(u) (uu),inner sep=2ex, yshift=2.4ex,
    minimum height=3.2cm] (ua) {};
    \node[above=-5.4ex of ua] {repository \\ aggregation};

    \node[draw,dashed,fit=(p) (pp) (ppp) (pppp) (ppppp), inner sep=2ex,
        yshift=2.4ex,
    minimum height=6.7cm] (pa) {};
    \node[above=-5.4ex of pa] {data exfiltration \\ functionalities};

    \node[draw,fit=(sa) (pa), inner sep=2.4ex, yshift=0.24cm,
    minimum height=7.8cm] (a) {};
    \node[above=-3.2ex of a] {BlackPOS};

\end{tikzpicture}
        \caption{Components and functionalities of BlackPOS. Yellow boxes are
            entry points of different functionalities.}
        \label{fig:blackposlogic}
    \end{center}
\end{figure}
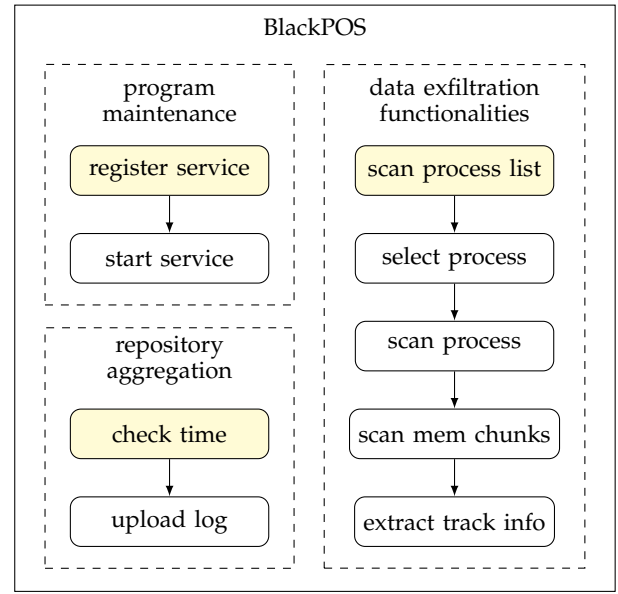

\subsection{Components and Functionalities of BlackPOS}

Belonging to the BlackPOS family, the malware discovered in the Target breach
is designed to infect Windows-based POS machines. The functionality of
BlacksPOS is not complicated and we present its components in
Fig.~\ref{fig:blackposlogic}.

When a POS terminal is infected, the malware registers itself as a Windows
service named ``POSWDS''. The service automatically starts with the operating
system, then \textit{i)} it scans a list of processes which could interact with
the card reader, and \textit{ii)} it communicates with a compromised server
(internal network repository) to upload retrieved credit card information.
Predefined rules apply for matching the sensitive processes, as well as
checking the time before sending obtained credit card numbers. Only during the
busy office hours in the daytime, the \textit{repository aggregation} function
could be enabled and the card information is sent to the internal repository.

Memory of target processes are read and analyzed in chunks, each of which is
10,000,000 bytes. BlackPOS uses a custom logic to search credit card numbers in
the memory trunks. It is believed that this method is more efficient and incurs
less overhead than generally used regular expressions~\cite{blackpos}.
Retrieved credit card information are encrypted and stored in file
``C:\textbackslash WINDOWS\textbackslash system 32\textbackslash winxml.dll''
and then periodically uploaded to the internal repository via NetBIOS and SMB
protocols.

\subsection{Design Features for Evading Detections}

BlackPOS evolves quickly during the past few years. The earliest versions of it
are discovered by McAfee in November 2011 as \texttt{PWS-FBOI} and
\texttt{BackDoor-FBPP}. They only contain the bare-bone logic for retrieving
and leaking sensitive information from individual machines~\cite{mcafeepos}.
However, the modern versions -- known to be used in the Target breach (2013)
and the Neiman Marcus breach (2013) -- are heavily customized for specific
internal networks and perform sophisticated behaviors to hide themselves from
common detection mechanisms. We detail multiple observed behaviors of BlackPOS
in the Target breach to illustrate how it is designed to circumvent detections.

\begin{itemize}

    \item \textit{Multi-phase data exfiltration.} Infected POS terminals do not
        send sensitive data to the external network directly. Instead, they
        gather data to a compromised internal server, which is used as a
        repository and one of the relies to reach the external
        network~\cite{targetsql}. The multi-phase data exfiltration scheme
        minimizes anomalous data flows across network boundaries.
        
    \item \textit{String obfuscation.} Critical strings in the malware
        executables are obfuscated to evade signature-based anti-virus
        detection~\cite{mcafeepos}. The strings include critical process names
        for scanning and NetBIOS commands for uploading data to the internal
        repository.

    \item \textit{Self-destructive code.} The malware avoids unnecessary
        infections to minimize its exposure. It destroys/deletes itself if the
        infected environment is not within its targets~\cite{battleblackpos}.
        This behavior reduces the risk of being detected in an unfamiliar
        environment.
         
    \item \textit{Data encryption.} The retrieved credit card information is
        encrypted in the file ``Winxml.dll'' in each POS terminal before it is
        sent to the internal repository. The encryption guarantees that no
        credit card numbers are sent in plaintext, which hides the leak from
        traditional data loss prevention (DLP) systems.
        
    \item \textit{Constrained communication.} Communications in the internal
        network are programed during office hours of the day~\cite{blackpos}.
        Busy office hour traffic helps hide anomalous communications between
        infected POS terminals and the compromised internal repository.

    \item \textit{Customized attack vector.} Internal IP addresses and login
        credentials of compromised servers are hardcoded in the malware. It
        indicates the malware author is aware of the internal network. The
        countermeasures against detections are deliberately designed along with
        the data exfiltration process.
        
\end{itemize}

\subsection{Malware Development and Marketing}

The Target breach attracts considerable attention to BlackPOS and similar POS
malware, e.g., vSkimmer~\cite{vskimmer} and Dexter~\cite{dexter}. Several
investigations have been performed to disclose the development and marketing of
these pieces of malware. Terrogence web intelligence company tracked the sales
of the malware on underground markets and pointed out BlackPOS was first posted
for sale in February 2013~\cite{posmalwaredates}. Cybercrime intelligence firm
IntelCrawler indicated Rinat Shibaev, a 17-year-old boy, and Rinat Shabayev, a
23-year-old Russian man, are the principle developers of
BlackPOS~\cite{teenager}. Andrew Komarov, CEO of the company, also hinted that
6 more retailer breaches are linked to BlackPOS~\cite{sixmore}. iSIGHT
Partners, working with United States Secret Service, investigated the POS
malware market and concluded a growing demand for such malware since 2010. FBI
tracked about 20 data breach attacks in recent years and warned retailers about
this increasing threat~\cite{FBIreport}.

\section{Prosecution of Data Breaches}
\label{sec:law}

The Target data breach is still under investigation and there is no arrest
known to the public. Tracking down data breach perpetrators is notoriously
difficult, because the criminals usually operate across the world to set
barriers for investigation and prosecution in terms of various laws and complex
treaties among countries. In this section, we discuss \textit{i)} the laws that
apply to cybercrimes, especially data breaches, \textit{ii)} the difficulties
in data breach discovery and prosecution, and \textit{iii)} a precedent of
investigation and sentence in the TJX breach case happened in 2007.

\subsection{Cybercrime Law and Regulations}

The federal Computer Fraud and Abuse Act (CFAA) is the most applicable
cybercrime law that applies to the Target breach itself. Other laws against
theft and misuse of the wires apply, as well as specific laws prohibiting the
sale of credit cards and identity theft~\cite{cybercrimelaws}.
Under the CFAA, unauthorized access to a computer engaged in interstate
commerce, which causes damage over \$5,000, is a crime punishable by 5 to 10
years in prison and up to \$250,000 damages, per offense. Subsequent violations
increase the potential penalty, and there are different provisions and
penalties for unauthorized access to government or financial computers. The
Federal Bureau of Investigation leads investigations and cases are prosecuted
by the Department of Justice Computer Crimes and Intellectual Property
Division. 

\subsection{Barriers to Data Breach Investigation}

Businesses, for a long time, declined to publically disclose a data breach in
fear that the information would hurt their reputation in the eyes of customers
and investors would. Today, 47 states have data breach notification laws.
Although not uniform, these laws generally require a business to report a data
breach to affected customers when personally identifiable information has been
lost. The requirement to report a data breach can aid law enforcement in
tracking down the criminals, and arguably is an incentive for businesses to
increase their security.

In data breach plots, attackers usually hide their identities carefully using
relays across the world in both the penetration phase (hacking into the system)
and the exfiltration phase (leaking the data out). The international relays
pose significant challenges for investigation and prosecution. In the Target
breach case, two drop sites are found in Miami and Brazil, and the final
aggregation server where all data is sent is discovered in Russia. There is no
guarantee that all involved countries take the same level of effort as the
United States to help investigate the incident. Each country is affected
differently by the breach, let alone the complicated relations mixed with
cooperation and divergences among them.

In addition, if cyber criminals are from outside the United States then an
arrest requires extradition from the foreign country. In order to extradite for
prosecution, the United States and the country must be signatories to a treaty
agreeing to such cooperation. Many countries in Asia, Africa and the Middle
East do not have treaties with the United States. Even with a treaty,
extradition involves a complicated process.

\subsection{TJX Breach and the Sentence}
\label{tjxbreach}

Before the Target data breach, 45.6 million credit card numbers and PINs were
stolen in the TJX data breach~\cite{tjxnumbers}. The breach was fully
investigated and the criminals were prosecuted and sentenced. The case sets a
record for credit card breach as well as the stiffest sentence for a
cybercrime. We describe details of the investigation from both technical and
legal aspects.

Albert Gonzalez, an American hacker, plotted the TJX data breach from July 2005
to January 2007. In addition to the TJX case, he was also charged with data
breaches in BJ's Wholesale Club, Boston Market, Barnes \& Noble, Sports
Authority, Forever 21, DSW and OfficeMax~\cite{tjxrelatedcomp}.

All aforementioned data breaches done by Gonzalez were carried out with similar
schemes. Taking the TJX case as an example, Gonzalez started with war-driving
along Route No. 1 in Miami to discover vulnerable retailer's hotspots. With the
help of his accomplices -- especially Stephen Watt, the author of the sniffer
used in the data breaches -- Gonzalez employed delicate SQL injections to gain
access to the database and to install the sniffer software into the servers.
Credit cards information was sniffed using ARP spoofing techniques and was
uploaded onto two foreign servers leased by Gonzalez in Latvia and Ukraine.

After obtaining the credit card information, Gonzalez sold the credit card
numbers and PINs to a Ukrainian card seller Maksym Yastremskiy. Yastremskiy
paid Gonzalez totaling \$400,000 through 20 electronic funds transfers via
\textit{e-gold} during 2006~\cite{gonzindictment}. He peddled the stolen credit
card information to other card sellers in the underground market. In 2007,
Yastremskiy was arrested on a separate charge, i.e., hacking into 12 banks in
Turkey.

In May 2008, Gonzalez was apprehended with \$1.1 million cash, a 2006 BMW, a
diamond and other assets.
He schemed to earn \$15 million from a series of data breaches, according to
his chat logs found by the government. He worked as an informant for the U.S.
Secret Service before he was arrested.


Gonzalez was sentenced on March 25th and 26th, 2010 for the TJX case and the
Heartland Payment Systems case, respectively. U.S. District Judge Patti Saris
sentenced Gonzalez to 20 years in prison,
and U.S. District Court Judge Douglas P. Woodlock sentenced Gonzalez to 20
years for the Heartland Payment Systems case. According to the negotiation
between Gonzalez and the government, the sentences run
concurrently~\cite{gonzalezmemo} and Gonzalez would be imprisoned for a total
of 20 years, which has reached record high on cybercrime~\cite{tjxhacker}.

\section{Lessons Learned Toward Better and More Effective Security Solutions}
\label{sec:companysolutions}

As we discussed in Section~\ref{sec:incident:security}, there are several
mistakes made by Target in the incident, including \textit{i)} ignoring
critical security alerts, \textit{ii)} improper segmentation of its network and
\textit{iii)} insecure point of sale data handling. In this section, we analyze
these three points in details and propose better design and more effective
practices for developing and deploying security solutions.

\subsection{Enforcing Payment System Integrity}

\begin{figure*}
    \begin{center}
        \includegraphics[width=6.7in]{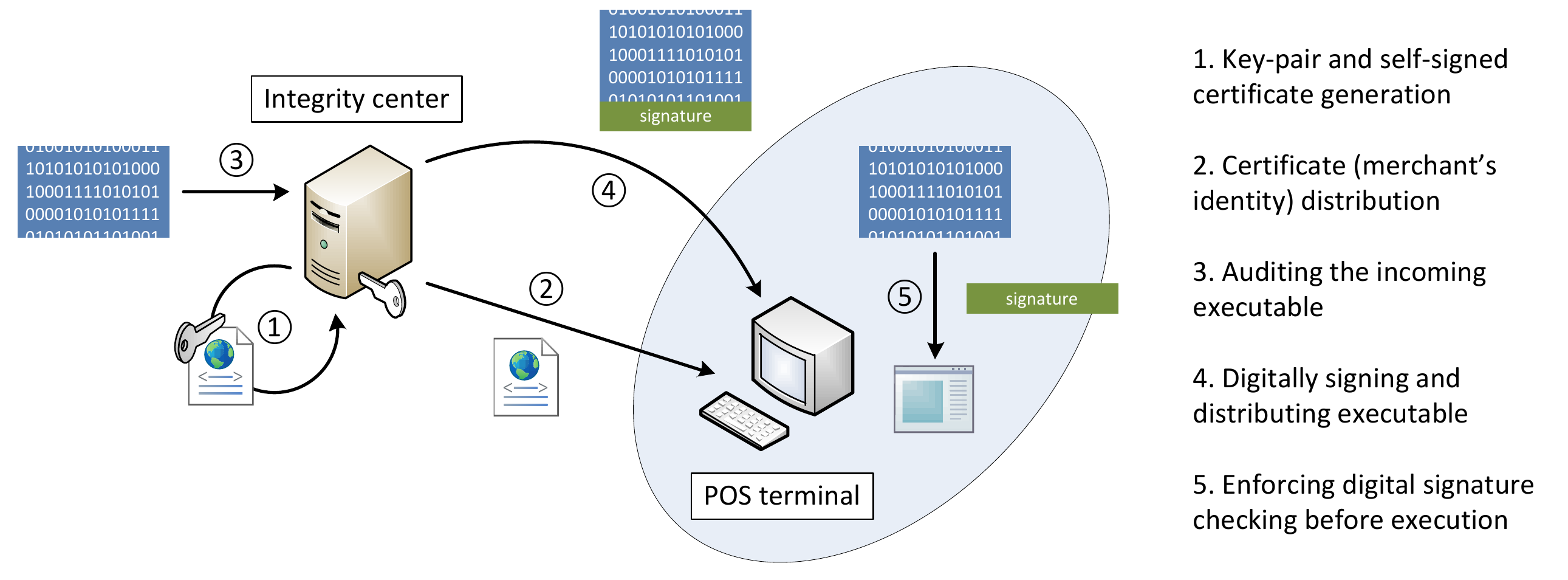}
        \caption{Our POS code update protocol with enhanced code integrity and authenticity verification.}
        \label{fig:posintegrity}
    \end{center}
\end{figure*}

In the Target breach, BlackPOS was installed on Target's point of sale
terminals, and the integrity of POS systems was compromised. This key step for
data breach can be prevented by enforcing the integrity of point of sale
terminals. Therefore, we provide a practical scheme using digital
signatures and certificates for ensuring the integrity of operating systems on
point of sales.

The workflow of our POS integrity scheme is shown in
Fig.~\ref{fig:posintegrity}. Our key idea is to allow only trusted executables
running on POS machines. An executable is trusted if it is verified/audited and
digitally signed by the merchant, i.e., Target Corp. 

Executable verification techniques such as digital signature for executables
are known for a long time, and many modern operating systems provide utilities
toward the goal, e.g., Microsoft Authenticode~\cite{authenticode}. However, the
execution policy is usually difficult to be enforced on a normal consumer's
computer because there are a variety of software providers on the Internet.
Users may install software or run programs from providers whose identify cannot
be verified. Public key infrastructure (PKI) helps relieve the issue, but it
does not completely solve it due to the complexity introduced by the variety of
software providers.

However, this approach is useful and practical in the dedicated environment
where \textit{i)} POS terminals are specifically used for processing
transaction and \textit{ii)} they are possessed and controlled by the merchant,
e.g., Target Corp. The first property ensures the software or programs running
on POS terminals are limited and feasible to be audited. The second property
guarantees one centralized \textit{integrity center} auditing and signing all
executables can be created.

There are two players, \textit{integrity center} and \textit{POS
terminal} and 5 steps in our integrity enforcement scheme. The integrity center
has four tasks: \textit{i)} key generation, \textit{ii)} key distribution,
\textit{iii)} file auditing and \textit{iv)} file signing. The POS terminal is
hardened by a policy that only binaries signed by the merchant can execute.
The five-step-protocol is:

\begin{enumerate}[~~~1.]
        
    \item The integrity center generates a public-private
        key pair $\langle pk, sk \rangle$ and creates a self-signed certificate
        $\textit{Cert}$ containing $pk$.
        
    \item The integrity center distributes $\textit{Cert}$ to every POS terminal in the
        company. $\textit{Cert}$ is placed in the root certificate list at each
        terminal.
        
    \item The integrity center audits every binary that needs to be executed on
        POS, e.g., programs, installers, system patches, etc. and signs the
        binary with $sk$ (encrypting the hash of the binary with $sk$).
        
    \item The signed binary is sent over the merchant network to POS terminals.
        
    \item The POS terminal checks the binary signature using $pk$ in $\textit{Cert}$
        (encrypting the signature with $pk$ to verify whether it is the hash of
        the binary) and executes only the ones correctly signed with $sk$.

\end{enumerate}

Adopting our payment system integrity enforcement protocol, merchants can achieve
the following two security goals in their system.

\begin{itemize}

    \item \textit{system integrity:} only trusted programs are allowed to be
        executed or installed on the payment system, which excludes the
        possibility of malware infection on point of sale devices.

    \item \textit{program authenticity:} every program or piece of software
        should pass the test at \textit{integrity center} before it is executed
        on point of sale machines, which allows merchants to have the full
        control of the payment system functionalities.

\end{itemize}

\subsection{Developing Effective Security Alert Systems}

\begin{figure*}
    \begin{center}
        \includegraphics[width=7in]{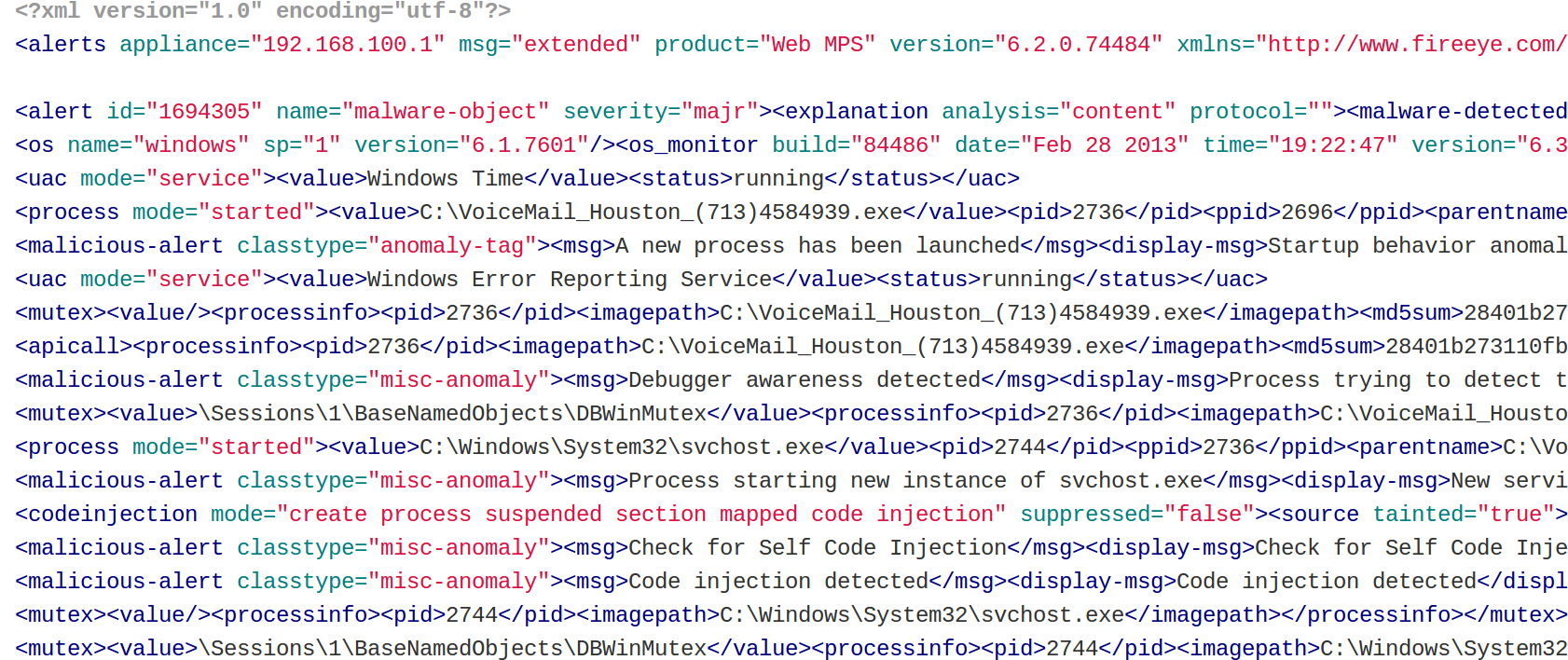}
        \caption{A FireEye alert in XML.}
        \label{fig:fireeye}
    \end{center}
\end{figure*}

Target had been warned multiple times by a malware detection tool produced by
FireEye Inc~\cite{bloomburg,reuters}. Unfortunately, the monitoring team in
Bangalore for Target Corp. took no actions in response to these alerts. They
also turned off the functionality that can automatically remove a detected
malware. These two serious mistakes hindered the detection of the leakage of
millions of credit card information. For large corporations, processing a large
number of security alerts produced by protection systems is challenging, if
possible at all. Many of these alerts are usually false alarms, which seasoned
security analysts learn to safely ignore. In this subsection, we first discuss
the design of FireEye alerts, and then explore new out-of-box design strategies
to improve the effectiveness of alerts.

\subsubsection{FireEye Alerts}

The raw data output from FireEye Threat Prevention Platform is in XML
structure. Fig.~\ref{fig:fireeye} shows a FireEye alert of a piece of
malware~\cite{fireeyeexample}. Basic information about the malware is provided,
such as \textit{type} and \textit{severity}. Anomalous behaviors of the malware
are tracked and listed in \textit{malicious-alert}. The
\textit{classtype=``anomaly-tag''} indicates that this alert is triggered
because of anomaly behavior detected.  The $msg$ and $display-msg$ briefly
describe the content of this alert.

In the Target case, FireEye alerted the administrators with type ``malware'',
which is commonly seen in large companies or organizations. However, no
sufficient detailed information was provided, e.g., the name of the malware or
the data exfiltration behavior of the malware. Since the BlackPOS software,
which extracts and steals sensitive financial information, is regarded as a
zero-day malware and few administrators have experience dealing with it, the
alerts were ignored~\cite{reuters}.

\subsubsection{Security Alert Design}

Security alert systems are at the front line of cyber defense. They represent
the first opportunity to detect, prevent, and stop attacks. Because human
analysts are error-prone and tend be undertrained, making alert systems more
usable and intelligent is critical. 

The needs for designing effective security warnings have been studied.
Sunshine et al. studied the effectiveness of SSL warning~\cite{sslwarining}.
Akhawe and Felt investigated the browser warnings including malware, phishing
and SSL warnings~\cite{browerwarning}. Modic and Anderson proposed to adopt
social-psychological techniques to increase the compliment for the warnings
~\cite{betterwarning}. 

Our thesis advocated in this paper on warnings differs from the existing
security alert research. We consider the security protection needs for large
companies and corporations that produce hundreds of alerts on daily basis. In
these scenarios, the alert systems need to handle and differentiate
warnings with a varying degrees of urgency. 


We argue that the design of alert systems needs to be adaptive and intelligent,
beyond simply sending a list of alerts. Specifically, we propose two design
strategies for security alerts: 

\begin{itemize}
        
    \item \textit{Adaptive warning strength.} Existing security systems provide
        severity information along with each alert, but there is no guarantee
        that important alerts are not ignored by administrators. Thus, we
        propose two methods to strengthen the efficiency of alerts:

        \begin{itemize}
           
            \item Raise the severity level of an alert when it is not handled
                within a limited amount of time. The purpose is to force
                security analysts to take actions toward severe alerts.
                This method is not applicable to all alerts,
                especially the less severe ones. Otherwise, it requires the
                administrators to address all alerts in the end, which may not
                be practical.
                 
            \item Besides color, font size, length of the alert bar, the system
                can raise alerts in different forms. For example, popping up
                flashing messages for the most critical alerts, emailing
                different level of alerts to different group of people.
                
                The multiform alarming method utilizes different ways to
                interact with different people. It also informs people who are
                not directly in charge of the issue to remind security analysts
                if issues are not solved for a long time.

        \end{itemize}

    \item \textit{Mining and presenting connections among alerts.} One drawback
        of existing detection solution is the lack of ability to correlate
        alert events. Some alert events could belong to a single attack vector
        and happen in sequence.  Sophisticated modern attacks are usually well
        planned and realized in steps. An alert may be triggered for each step,
        e.g., malware injected, file transmission. Connecting these alerts can
        reveal a grander scheme of the plot.

        One approach to connect alerts is to analyze the consequences of each
        malicious event. The consequences can be used to bridge alerts, connecting 
        multiple alerts in different types to a plot. If the collection alerts indicate 
        potential grander data breach, then sever alerts should be raised.

\end{itemize}

\subsection{Controlling Information Flow with Network Segmentation}

Target failed to segment its sensitive assets from normal network portions,
which allows an attacker to escalate the intrusion if he/she attacks from the
inside. We explain the severity of this issue.


The most common strategies used in network architecture are techniques based on
building a strong exterior, so that only those a system can trust can get
inside. Because the only people inside are those who can be trusted, security
on the internal network is either low or none existent. An example of this goes back to the Target breach.
Once the attacker obtained the security credentials from Fazio Mechanical
Services, they then had the ability to gain access into Target's network. 
From there they compromised three FTP servers and installed malware on
many point of sales devices~\cite{targetfireeye}.

The security principle advocated in a zero trust network~\cite{forrester} is 
simply don't trust anyone. This means all traffic is
identified, authorized, and monitored. This makes all parts of the network
secure regardless of location. In addition, virtual LANs cannot provide much
security defense, because they cannot stop an intruder from gaining access into
other portions of the network. The Target incidence demonstrates that virtual
LANs, especially when not configured properly, is ineffective against the
criminals.

While the zero trust strategy protects from outsider attacks it also protects
against insider attacks because all traffic is monitored and analyzed. If a
member of a network does something unusual, e.g. deletes several entries in a
database that he or she usually does not access, the network administrators can
detect the change in behaviors. However, this strategy has the trade-off for
usability because monitoring all traffic leads to extremely huge computation
power. And it is not convenient for practical usage in large scale networks.

\section{Credit Card Security and Best Practices for customers}
\label{sec:customersolutions}

Target and other breach incidents, e.g., Neiman Marcus, Sally's Beauty and PF
Chang's, suggests that there is a high risk at current credit card regulation
and technology. In this section, we discuss the issues in credit card
regulation as well as advantages and problems of new technologies for securing
credit card transactions.

\subsection{Credit Card Administration and Regulation}

Payment card security is self-regulated by the contract between the merchant
and the card company. Major credit card companies require compliance with the
Payment Card Industry Data Security Standard (PCI-DSS)~\cite{pcidss}. The
description of Target’s security, such as weak password at the POS, would not
seem to meet many of the standards, thus drawing attention to whether the
private contract self-regulation framework is effective.

\subsection{EMV: Toward a More Secure Payment System}

EMV (Europay, MasterCard, and Visa) payment system is the major technology
developed to address the security issue in credit cards~\cite{anderson2014emv}.
The EMV system adds a temper-resistant chip to a credit card. The chip stores
confidential account information and provides on-chip cryptographic
computations such as encryption and digital signing. The system works by
authenticating through the chip and identifying the user by either a signature
or a pin; this is where the system gets its name of \textit{chip and pin}. The
difference between the EMV system and the traditional magnetic strip system is
that the data on the card's chip is encrypted. Therefore it would be much more
difficult for an attacker to commit fraud.

However, a flaw is found in the design of the transaction protocol, which
makes EMV ineffective. A no-pin attack is the scenario where the criminal has
the card but not the pin. Murdech et al. show that by using an electric device
between the card and the terminal, the terminal can be tricked into believing
that the criminal has the correct pin even though he
doesn't~\cite{chippinbroken}.

Another vulnerability is found in Point of Sale terminals. A POS terminal in
the EMV system is assumed temper-resistant, meaning that no one can open the
POS box and read/write to the internal circuits. Unfortunately, real
EMV-equipped POS terminals can be tempered and authentication codes can be
obtained and used at a later time on the same terminal to make additional
transactions. Several criminals in Spain made use of this vulnerability as
well as a vulnerability found in ATM random number generators. An ATM may
generate predictable random numbers which gives criminals temporary access to
credit card spending if the number is guessed correctly.

Besides the two vulnerabilities, the most severe flaw in the EMV system is the
card not present (such as when a purchase is made online) fraud abbreviated as
CNP fraud. CNP fraud now accounts for over fifty percent of fraud in the United
Kingdom where the EMV system is currently being used~\cite{acquirer}. The EMV
system is designed for securing card-present transactions, and it has nothing
in place to prevent CNP fraud from occurring, which is why criminals are now
using CNP fraud as their go to choice in fraudulent purchases. 

\subsection{Tokenization and Best Practices for Customers}

Tokenization is a payment technology to minimize credit card information by
merchants during transactions. In this section, we describe the technology and
explain why it helps protect personal account information. We give customers
best practices to hide their credit card information when shopping.

With tokenization
a customer asks an acquirer to act between he/she and the merchant. The
acquirer \textit{i)} takes the customer's credit card information $c$,
\textit{ii)} generates a one-time token $t$ based on $c$ ($t$ is independent of
$c$), and \textit{iii)} sends $t$ to the merchant to process the transaction.
$t$ is bound to the merchant and can be nullified after the transaction.

There are two approaches to utilize an acquirer. One is to pay through acquirer
systems. Available systems include PayPal, Amazon payment, Google wallet and
Apple pay. For example, customer \textit{Alice} wants to buy an item in Amazon
market from seller \textit{Bob}. \textit{Alice} can pay through the Amazon
payment system and \textit{Bob} only receives a token to process the
transaction. Google wallet and Apple pay extend the service from online
shopping to in-store purchases. They also provide contactless feature through
near field communication (NFC), so that \textit{Alice} does not need to swipe a
card to authorize a purchase.

The second approach to utilize an acquirer is to generate a one-time credit
card number at acquirer banks. Bank of America and Citibank provide such
service, namely \textit{ShopSafe} and \textit{virtual card number},
respectively. PayPal used to have a similar service and Discover terminated its
virtual card service on March 16, 2014.




        



\section{Conclusion}
\label{sec:conclusion}

There is no silver bullet in cyber space against data breaches.
With the increasing amount of data leak incidents in recent years, it is
important to analyze the weak points in our systems, techniques and
legislations and to seek solutions to the issue. In this paper, we presented a
comprehensive analysis of the Target data breach and related incidents, such as
the TJX breach. We described several security guidelines to enhance security in
merchants' systems. We presented the state-of-the-art credit card security
techniques, and gave customers best practices to hide card information during
purchase transactions.


\section*{Acknowledgments}

This work has been supported in part by NSF grant CAREER CNS-0953638 and ARO
grant YIP W911NF-14-1-0535.

\bibliographystyle{IEEEtran}
\bibliography{target}

\end{document}